\documentclass[11pt,preprintnumbers,amsmath,amssymb]{revtex4-1}

\usepackage{graphicx}              % Include figure files
\usepackage{dcolumn}               % Align table columns on decimal point
\usepackage{longtable}             % helps with long table options
\usepackage{bm}                    % bold math
\usepackage{afterpage}
\usepackage{color}
\usepackage{dsfont}
\usepackage{framed}
\usepackage{algorithm}
\usepackage[noend]{algpseudocode}

\def\beq{\begin{eqnarray}}
\def\eeq{\end{eqnarray}}
\def\beqq{\begin{eqnarray*} \color{blue} }
\def\eeqq{\end{eqnarray*}}
\def\OpOne{\hat{\mathbf{1}}}
\def\rvec{{\mathbf r}}

\begin{document}

\title{Efficient heat-bath sampling in Fock space}

\author{Adam Holmes$^1$}
\email{aah95@cornell.edu}
\author{Hitesh J. Changlani$^{2}$}

\author{C. J. Umrigar$^1$}
\email{CyrusUmrigar@cornell.edu}
\affiliation{
  $^1$Laboratory of Atomic and Solid State Physics, Cornell University, Ithaca, New York 14853, USA\\
  $^2$Department of Physics, University of Illinois at Urbana-Champaign, Urbana, Illinois 61801, USA\\
}

\begin{abstract}
We introduce an algorithm for sampling many-body quantum states in Fock space.
The algorithm efficiently samples states
with probability approximately proportional to an arbitrary function of the
second-quantized Hamiltonian matrix element connecting the sampled state to the current state.
We apply the new sampling algorithm to the recently-developed
Semistochastic
Full Configuration Interaction Quantum Monte Carlo method (S-FCIQMC),
a semistochastic implementation of the power method for projecting out
the ground state energy in a basis of Slater determinants.
The heat-bath sampling requires modest additional computational time and memory compared to uniform sampling
but results in newly-spawned weights that are approximately of the same magnitude, thereby
greatly improving the efficiency of projection.
A comparison in efficiency between uniform
and approximate heat-bath sampling is performed on the all-electron
nitrogen dimer at equilibrium in Dunning's cc-pV$X$Z basis sets with $X\in\left\{ {\rm D},{\rm T},{\rm Q},5\right\} $,
demonstrating a large gain in efficiency that increases with basis
set size.
In addition, a comparison in efficiency is performed on
three all-electron first-row dimers, B$_2$, N$_2$, and F$_2$, in a cc-pVQZ basis,
demonstrating that the gain in efficiency compared to uniform sampling
also increases dramatically with the number of electrons.
\end{abstract}
\maketitle

\section{Introduction}

Methods for finding approximate solutions to the quantum many-body problem often
work in Fock space, spanned by a discrete basis composed of products of
$N$ single-particle orbitals chosen from a set of $M$ orbitals, where $N$ is the number of particles.
These products are symmetrized or antisymmetrized for bosonic
or fermionic particles, respectively.
However, the number of states in Fock space scales exponentially in $M$,
and even the size of the sector of Fock space with constant particle number $N$ sector scales combinatorially in $M$ and $N$,
so deterministic calculations are limited to small $M$ and $N$.

When the number of states is too large for deterministic methods (approximately $10^{10}$),
Monte Carlo methods provide a valuable alternative.
For example, the Full Configuration Interaction Quantum Monte Carlo (FCIQMC)
method~\cite{BooThoAla-JCP-09,CleBooAla-JCP-10,SheGruBooKreAla-PRB-12,BooSmaAla-MP-14},
and its semistochastic extension, the S-FCIQMC method~\cite{PetHolChaNigUmr-PRL-12}
have been used to calculate almost exact energies in far larger state spaces.
Another example is the Monte Carlo Configuration Interaction (MCCI) method~\cite{greer1998monte},
in which a variational wavefunction is calculated from a selected CI expansion generated by a Monte Carlo sampling procedure.

In both of these methods, Slater determinants are sampled \emph{uniformly}
from the set of determinants connected to a reference by the Hamiltonian.
While uniform sampling has the advantage of being easy to implement,
it is far from optimal, since some of the determinants connected to the
reference are much more important than others.
The definition of ``importance" here depends on the method used.
In S-FCIQMC, starting from an initial state, the importance of a final state is
the magnitude of the Hamiltonian matrix element connecting it to the initial state:
if moves are proposed with probabilities proportional to this importance function,
the weight of the final state is independent of which of the possible final states is chosen.
In MCCI, the importance of a state is the magnitude of the 2nd-order perturbation theory estimate
of energy lowering achieved by including this state in the expansion.
Since the importance of the states can range over many orders of magnitude, large
gains in efficiency can be gained by sampling states in proportion to their importance.
However, the na\"{\i}ve approach of computing the relative probabilities of all the connected states
and then normalizing by dividing by their sum is prohibitively expensive because %results in a loss of efficiency because
the number of connected states can be large.
For example, for a quantum chemistry Hamiltonian, it scales as $\mathcal{O}\left(N^2 M^2\right)$.

Motivated by the above, this paper addresses the following problem:
Given an initial configuration in Fock space, how can one \emph{efficiently} sample new configurations
with probability approximately proportional to
a function of the Hamiltonian matrix element connecting the new configuration to the initial one?
By `efficient', we mean that the time complexity of one sampling event scales as $\mathcal{O}\left(N\right)$,
the same time complexity as that of computing one Hamiltonian matrix element corresponding to a single excitation in a molecular system.
We also limit our storage requirements to $\mathcal{O}\left(M^k\right)$, where $k$ is the
maximum number of creation and annihilation operators in any off-diagonal term in the Hamiltonian $-$ in other words, the same
scaling as that required to store the integrals needed for computing the Hamiltonian.
We take advantage of the fact that $k$ is typically small; e.g., it is 4 for the quantum chemistry Hamiltonian
and for the Hubbard model in momentum space, and it is 2 for the Hubbard model in real space.
We demonstrate the efficiency gains achieved when the method is used in S-FCIQMC applied to the
quantum chemistry Hamiltonian.

This paper is organized as follows.  Section~\ref{background} has an overview of the S-FCIQMC method
and its application to the quantum chemistry Hamiltonian.
In Section~\ref{heatbath} the approximate heat-bath
sampling method is discussed within the context of S-FCIQMC.
In Section~\ref{results}
uniform and heat-bath sampling methods are compared for S-FCIQMC calculations on the
nitrogen dimer at equilibrium in several different basis sets, demonstrating that heatbath sampling
is more efficient than uniform sampling by a factor that increases rapidly with basis size.
Similar calculations are also
performed on the boron dimer and the fluoride dimer,
demonstrating that at fixed number of orbitals, the gain in efficiency increases even
more rapidly with electron number.

\section{Background}
\label{background}

\subsection{Quantum chemistry Hamiltonian}

The second-quantized nonrelativistic electronic Hamiltonian,
in the Born-Oppenheimer approximation is,
\beq
\hat{H} &=&\sum_{pr} f_{rp} a_{r}^{\dagger}a_{p}+ \frac{1}{2} \sum_{pqrs}g_{rspq}a_{r}^{\dagger}a_{s}^{\dagger}a_{q}a_{p}+h_{{\rm nuc}}, % phys notation
\label{eq:chemham}
\eeq
where $a_{p}^{\dagger} (a_{p})$ is the usual electron creation (destruction) operator with the
indices $\{p,q,r,s\}$ incorporating both spatial and spin degrees of freedom.
The terms entering the expression of the Hamiltonian are,
\beq
f_{rp} & = & \int\phi_{r}^{*}\left(x\right)\left(-\frac{1}{2}\nabla^{2}-\sum_{I}\frac{Z_{I}}{\left| \rvec_{I} - \rvec \right| }\right)\phi_{p}\left(x\right)dx
\eeq
the 1-center integrals with $\phi_{p}(x)$ denoting spin-orbitals, $x$ the combined spatial ($\rvec$) and spin coordinates of the electrons, and $Z_{I}$ and $\rvec_I$
the atomic number and spatial coordinates of nucleus $I$,
\beq
g_{rspq} & = & \int\phi_{r}^{*}\left(x_{1}\right)\phi_{s}^{*}\left(x_{2}\right)\frac{1}{\left| \rvec_{1} -\rvec_{2} \right|}\phi_{p}\left(x_{1}\right)\phi_{q}\left(x_{2}\right)dx_{1}dx_{2}
\eeq
the 2-center integrals with an index-ordering convention according to the
physicist notation~\cite{physicist_chemist_notation},
and
\beq
h_{{\rm nuc}} & = & \sum_{I<J}\frac{Z_{I}Z_{J}}{\left| \rvec_{I} -\rvec_{J} \right|}
\eeq
the nuclear-nuclear repulsion.
Orthogonal orbitals are used, and in the absence of a magnetic field
can be chosen to be real in which case $f_{rp}=f_{pr}$ and
 $g_{rspq}=g_{psrq}=g_{rqps}=g_{pqrs}=g_{srqp}=g_{qrsp}=g_{spqr}=g_{qpsr}$. % phys notation
Hence the storage required for the 1-center integrals is ${\cal O} (N^2/2)$ and
that for the 2-center integrals is ${\cal O} (N^4/8)$.

Since the Hamiltonian in Eq.~\eqref{eq:chemham} contains only 1-electron and 2-electron terms,
it has nonzero matrix elements between two determinants only if they differ by no more than two spin-orbitals.
Classifying the matrix elements by the number of spin-orbitals
that the initial ($i$) and final states ($f$) differ in,
\beq
H_{\rm diag} &=& \sum_{p\in \rm occ.} f_{pp}   + \frac{1}{2}\sum_{p,q \in \rm occ.}(g_{ppqq}-g_{pqqp}),
\label{diag} \\
H(r \leftarrow p) &=& \Gamma^{i}_{rp} \left( f_{rp}  + \sum_{q\in \rm occ.}(g_{rpqq}-g_{rqqp})\right),
\label{single_excit} \\
H(r s \leftarrow p q ) &=& \Gamma^{i}_{rp}\Gamma^{f}_{sq} \bigg( g_{rspq}-g_{rsqp}\bigg),
\label{double_excit}
\eeq
where $\Gamma^{i}_{rp} = (-1)^n$, $n$ being the number of occupied
spin-orbitals between $p$ and $r$ in state $i$.

Note that the magnitudes of the matrix elements for double excitations depend only on
the four spin-orbitals whose occupations are changing, though the sign depends on the other
occupied spin-orbitals as well.  In contrast the magnitudes of the diagonal
and the single excitation matrix elements depend on all the occupied spin-orbitals.

\subsection{S-FCIQMC overview}
We now briefly review the S-FCIQMC method~\cite{PetHolChaNigUmr-PRL-12} which is a
modification of the FCIQMC method~\cite{BooThoAla-JCP-09,CleBooAla-JCP-10,SheGruBooKreAla-PRB-12,BooSmaAla-MP-14}.
There are three main differences between the S-FCIQMC method and the original FCIQMC method, namely,
walkers have real rather than integer weights, the computation of the mixed estimator
of the energy is done using a multideterminantal trial wavefunction rather than the
Hartree-Fock determinant and the part of the projection that involves the most
important determinants is done deterministically rather than stochastically.

In common with all QMC methods, the S-FCIQMC method employs a ``projector" operator,
which is a function of the Hamiltonian such that the ground state of the Hamiltonian
is the dominant state (state with largest absolute eigenvalue) of the projector.
Repeated application of the projector to an arbitrary state that is not orthogonal
to the ground state results in projecting onto the ground state.
The S-FCIQMC method employs the linear projector,
\beq
\hat{G} &=& \OpOne+\tau\left(E_{\rm T}-\hat{H}\right),
\eeq
where $E_{\rm T}$ is an estimate of the ground state energy and $i,j$ denote Fock space states.
The time step, $\tau$, must be smaller than $2/(E_{\rm max}-E_{\rm min})$,
where $E_{\rm max}$ and $E_{\rm min}$ are the extremal eigenvalues of $\hat{H}$, in order
for the ground state of $\hat{H}$ to be the dominant state of $\hat{G}$.
In order to avoid negative diagonal matrix elements in $\hat{G}$, $\tau$, must be smaller than $1/(E_{\rm max}-E_{\rm min})$.
In practice, $\tau$ is chosen to roughly minimize the statistical error for given
computer time, and this optimal value is yet smaller.

Since the total number of states is much larger than the number of states that can be
stored on a computer, it is necessary to do at least part of the projection stochastically.
Before the run, a subset of the determinants is selected to be dubbed the {\it deterministic space}.
Projection between pairs of determinants that are both in the deterministic
space is performed deterministically, using a sparse matrix-vector multiplication.
However, projection between a pair of determinants at least one of which is outside the
deterministic space is performed stochastically, as follows.

At any Monte Carlo (MC) step, $t$, the current state is represented as a sparse
linear combination of ``occupied" states,
\beq
\left|\psi_0^{t}\right> &=& \sum_{i\in {\rm occ.}}w_{i}^{t}\left|i\right\rangle ,
\eeq
where $w_{i}^{t}$ is the weight on state $\left|i\right\rangle$ at step $t$.
The state evolves from MC step $t$ to step $t+1$ according to
\begin{eqnarray}
w_{j}^{t+1} & = & \sum_{i}\hat{G}_{ji}w_{i}^{t} \nonumber\\
 & = & \sum_{i}\left[\delta_{ji}+\tau\left(E_{\rm T}\delta_{ji} - \hat{H}_{ji}\right)\right]w_{i}^{t} \nonumber\\
 & = & \left(1+\tau\left(E_{\rm T} - \hat{H}_{ii}\right)\right)w_{i}^{t}-\tau\sum_{i\ne j}\hat{H}_{ji}w_{i}^{t}.
\end{eqnarray}
Note that the first of these terms is diagonal, and the second is
off-diagonal. In addition to the off-diagonal elements
connecting pairs of deterministic determinants,
all diagonal elements can be applied deterministically,
since they do not increase the density of the sparse representation.
However, the off-diagonal elements that do not connect pairs of
deterministic states are sampled stochastically as follows.

On each iteration, the weight $w_i^t$ on each Slater determinant $i$ is divided
up into an integer number of walkers $n_i^t=\max\left(1,\lfloor \left| w_i^t \right|\rceil\right)$, and each walker spawns a new
walker on determinant $j$ with probability $P_{ji}$
that receives a weight of
\beq
w_{j}^{\left(t+1\right)} &=&\frac{-\tau H_{ji}}{P_{ji}}\left(\frac{w_{i}^{t}}{n_i^t}\right).
\label{weight}
\eeq

To overcome the fermion sign problem, only determinants with absolute
weight greater than an \emph{initiator} threshold~\cite{CleBooAla-JCP-10} are allowed to create
weight on unoccupied determinants. The resulting bias in the energy
disappears in the limit of infinite walker number.
For efficiency reasons, walkers with weight less than a minimum weight
are combined into a smaller number of larger weight walkers in a
statistically unbiased fashion~\cite{PetHolChaNigUmr-PRL-12}.
In this work, the minimum weight is set to 0.5, and a graduated initiator is used:
the initiator threshold for a given determinant is equal to
the minimum number of moves a walker on that determinant has
taken since its last visit to the deterministic space.

An estimate of the ground state wavefunction could in principle be obtained by summing the
walker distributions over all Monte Carlo iterations over a long run, i.e.,
\beq
\left|\psi_0\right> \approx \sum_{t=1}^{N_{\rm MC}}\left|\psi_0^{t}\right>= \sum_{t=1}^{N_{\rm MC}} \sum_{i\in {\rm occ.}}w_{i}^{t}\left|i\right\rangle.
\eeq
However, the entire wavefunction $\left|\psi_0\right\rangle$ contains too many terms to be stored all at once;
instead, only the sparse walker distribution
$\left\{w_i^{t}\right\}$
at a given iteration can be stored.
Therefore, $E_{0}$ must be estimated in a way that depends
only on linear functions of the walker distributions.
This is accomplished using a mixed estimator,
\beq
E_{0}^{\rm mix}=\frac{\left\langle \psi_{0}\left|\hat{H}\right|\psi_{T}\right\rangle }{\left\langle \psi_{0}|\psi_{T}\right\rangle }
=\frac{\sum_{t=1}^{N_{\rm MC}}\sum_{i\in{\rm occ.}}w_i^{t}N_i}{\sum_{t=1}^{N_{\rm MC}}\sum_{i\in{\rm occ.}}w_i^{t}t_i},
\eeq
where
$\left|\psi_{T}\right\rangle=\sum_i t_i\left|i\right\rangle$
is a trial wavefunction that is chosen before the run, and
$N_i=\sum_j H_{ij} t_j$ can also be computed and stored before the run.
In the limit that either $\left|\psi_0\right\rangle$ or $\left|\psi_T\right\rangle$
is the exact ground state, the mixed energy is a zero-bias, zero-variance
estimator of $E_0$. Thus, we choose $\left|\psi_T\right\rangle$
to be a low-energy linear combination of determinants.

Before starting an S-FCIQMC run, two sets of important determinants must be selected:
the deterministic space and the determinants that make up the trial wavefunction
expansion (after selecting the determinants that make up $\left|\psi_T\right\rangle$, a Lanczos diagonalization
is then performed within that set of determinants to obtain the coefficients).
In both important subspaces, including more determinants improves the efficiency,
but the two subspaces have different storage constraints. The deterministic space
must be small enough that the full Hamiltonian within the deterministic space can be
stored (as an upper-triangular sparse matrix). The number of determinants in $\left|\psi_T\right\rangle$
is limited by the requirement that all of the mixed energy numerators $N_i=\sum_j H_{ij} t_j$
can be stored. This means that the deterministic space can typically be much larger than
the trial wavefunction expansion.

To select determinants for either subspace, the following iterative procedure is used.
Starting with the Hartree-Fock determinant, all connected determinants are obtained.
Second-order perturbation theory is then used to estimate the coefficient each
determinant would have if a Lanczos diagonalization were performed;
only the 10\% of the determinants with largest expected coefficients are retained.
A Lanczos diagonalization is performed in the resulting space.
Then, the determinants connected to the subset of determinants with the highest absolute coefficients
are obtained, and the procedure is repeated.
Finally, the determinant list is truncated to the required size, and
in the case of the trial wavefunction, a final Lanczos diagonalization is performed.

\subsection{Proposal of off-diagonal moves}

The computed energy does not depend on $P_{ji}$ in Eq.~\ref{weight},
provided that this is in fact the probability of proposing the move.
However, the statistical error does depend on $P_{ji}$.
Large weight fluctuations increase the statistical error in any Monte Carlo calculation,
and their detrimental effect is even more severe when there is a sign problem.
If the new states are chosen from the \emph{heat-bath} distribution, i.e., with probability proportional
to the magnitude of the corresponding matrix element, i.e.,
\beq
P_{ji} &=& \frac{\left|H_{ji}\right|}{\sum_{k}\left|H_{ki}\right|}.
\eeq
then their weights are independent of which state is chosen.

The difficulty with heat-bath sampling is that it is
prohibitively expensive to compute the full column sum $\sum_{k}\left|H_{ki}\right|$
every time an off-diagonal move is proposed, since the number of
off-diagonal elements is $\mathcal{O}\left(N^{2}M^{2}\right)$.
Hence, until now, most S-FCIQMC calculations have been performed using
an approximately uniform sampling of off-diagonal moves, which is
computationally simpler since it only involves the number of off-diagonal elements
$-$ which can be computed before the run $-$
rather than computing and summing them each iteration.

In this paper, we introduce an efficient new approach for sampling an approximate
heat-bath distribution, in which we factor the above proposal
probability into probabilities of selecting each electron and unoccupied spin-orbital
separately. These probabilities of individual steps can be computed and stored
before the run in order to reduce the sampling time to
$\mathcal{O}\left(N\right)$, which is the same scaling as that for computing one
single-excitation matrix element.
The algorithm is derived in the next section, and is summarized concisely in
Appendix~\ref{Algorithm}.

\section{Approximate heat-bath algorithm}
\label{heatbath}

The nonrelativistic quantum chemistry Hamiltonian contains off-diagonal terms corresponding
to single and double excitations only.
Double excitations are more numerous, but much simpler to deal with than single excitations.
While the number of double excitations from a given determinant is
$\mathcal{O}\left(N^2M^2\right)$,
the number of distinct values of double excitation matrix elements in the full Hamiltonian is only $\mathcal{O}\left(M^4\right)$,
and each double excitation matrix element takes only $\mathcal{O}\left(1\right)$ time to compute.
On the other hand, while the number of single excitations possible from a given determinant is only
$\mathcal{O}\left(NM\right)$,
the number of distinct values of single excitation matrix elements in the full Hamiltonian is combinatorially large,
and each matrix element takes $\mathcal{O}\left(N\right)$ time to compute,
since the matrix elements for exciting from a given spin-orbital include a sum over all other occupied spin-orbitals (see Eq.~\ref{single_excit}).
So, efficient computation of single-excitation matrix elements is difficult because
there are too many to store and they are expensive to compute on the fly.

Since double excitations account for most of the possible excitations from a given determinant,
and since they are so much easier to deal with efficiently, we start by constructing an approximate heat-bath algorithm
for double excitations, and then we incorporate single excitations into the algorithm.

\subsection{Sampling double excitations}

If single excitations are ignored, heat-bath sampling requires us to
sample a move of electrons from spin-orbitals $\left\{p,q\right\}$ to spin-orbitals
$\left\{r,s\right\}$ with probability
\beq
P\left(r s \leftarrow p q\right) &=& \frac{\left|H(r s \leftarrow p q)\right|}{\sum_{p'q'\in \rm occ.}\sum_{r's'\in \rm unocc.}\left|H(r' s' \leftarrow p' q')\right|},
\eeq
where the sum over $\left\{p',q'\right\}$ is over all occupied spin-orbitals, and the sum
over $\left\{r',s'\right\}$ is over all unoccupied spin-orbitals.

We can factor this probability into a four-step process as follows:
\begin{eqnarray}
P\left(r s \leftarrow p q\right) & = & \frac{\left|H(r s \leftarrow p q)\right|}{\sum_{p'q'}\sum_{r's'}\left|H(r' s' \leftarrow p' q')\right|}\nonumber \\
 & = & \left(\frac{\sum_{q'\in \rm occ.}\sum_{r's'\in \rm unocc.}\left|H(r' s' \leftarrow p q')\right|}{\sum_{p'q'\in \rm occ.}\sum_{r's'\in \rm unocc.}\left|H(r' s' \leftarrow p' q')\right|}\right)
\left(\frac{\sum_{r's'\in \rm unocc.}\left|H(r' s' \leftarrow p q)\right|}{\sum_{q'\in \rm occ.}\sum_{r's'\in \rm unocc.}\left|H(r' s' \leftarrow p q')\right|}\right)\nonumber \\
&&\times\left(\frac{\sum_{s'\in \rm unocc.}\left|H(r s' \leftarrow p q)\right|}{\sum_{r's'\in \rm unocc.}\left|H(r' s' \leftarrow p q)\right|}\right)
\left(\frac{\left|H(r s \leftarrow p q)\right|}{\sum_{s'\in \rm unocc.}\left|H(r s' \leftarrow p q)\right|}\right)\nonumber \\
 & = & P\left(p\right)P\left(q|p\right)P\left(r|p,q\right)P\left(s|p,q,r\right).
\end{eqnarray}

The conditional probabilities of selecting unoccupied spin-orbitals
for double excitation, $P\left(r|p,q\right)$ and $P\left(s|p,q,r\right)$,
will be different for each determinant, since they involve sums over
all the spin-orbitals that are currently unoccupied. However, we can approximate
them by summing over \emph{all} spin-orbitals, except for $p$ and $q$:
\beq
P\left(r|p,q\right)\approx\tilde{P}\left(r|p,q\right)\equiv\frac{\sum_{s'\notin\left\{ p,q\right\} }\left|H(r s' \leftarrow p q)\right|}{\sum_{rs\notin\left\{ p,q\right\} }\left|H(r' s' \leftarrow p q)\right|},\quad
P\left(s|p,q,r\right)\approx\tilde{P}\left(s|p,q,r\right)\equiv\frac{\left|H(r s \leftarrow p q)\right|}{\sum_{s'\notin\left\{ p,q\right\} }\left|H(r s' \leftarrow p q)\right|}.
\eeq
Spin-orbitals can be sampled from these distributions in ${\cal O}(1)$ time using the alias method~\cite{walker1977efficient},
as described in Appendix~\ref{Alias}.

A similar approximation can be made for the probability of selecting
the occupied spin-orbitals, $P\left(p\right)$ and $P\left(q|p\right)$. %$P\left(p,q\right)=P\left(p\right)P\left(q|p\right)$.
Defining
\beq
D_{pq} & \equiv & \sum_{r's'\notin\left\{ p,q\right\} }\left|H(r' s' \leftarrow p q)\right|\quad {\rm and}\quad
S_{p}\equiv\sum_{q'\ne p}D_{pq'},
\eeq
we can approximate $P\left(p\right)$ and $P\left(q|p\right)$ by
\beq
P\left(p\right)\approx\tilde{P}\left(p\right)\equiv\frac{S_{p}}{\sum_{p'\in \rm occ.}S_{p'}},\quad
P\left(q|p\right)\approx\tilde{P}\left(q|p\right)\equiv\frac{D_{pq}}{\sum_{q'\in \rm occ.}D_{pq'}}.
\eeq
Since $S_{p}$, $D_{pq}$, $\tilde{P}\left(r|p,q\right)$, and $\tilde{P}\left(s|p,q,r\right)$
all involve sums over \emph{all} spin-orbitals, they can be computed and stored once at the beginning of the run.
However, $\tilde{P}\left(p\right)$ and $\tilde{P}\left(q|p\right)$ involve sums over only the currently-occupied
spin-orbitals, and thus they must be computed on the fly (in $\mathcal{O}\left(N\right)$ time).

Note that approximating the heat-bath probabilities affects only the efficiency of the
calculation, not its exactness. Once the initiator bias has been extrapolated away,
the results are exact (i.e., they have a statistical error but
no bias), so long as the same probabilities are used for proposing moves and computing the weights.

\subsection{Sampling single excitations}

We now modify the double-excitation proposal algorithm above
to include the possibility of proposing a single excitation.
After selecting electrons in spin-orbitals $\left\{p,q\right\}$ and empty spin-orbital $r$, we
allow two possibilities:

\begin{enumerate}
\item Select a second unoccupied spin-orbital
$s$, generating the double excitation $\left(r s \leftarrow p q\right)$
(as described above), OR

\item Discard the already-selected occupied spin-orbital $q$ and
generate the single excitation $\left(r \leftarrow p\right)$.%, OR

\end{enumerate}

The heat-bath distribution
requires us to choose between a single and a double excitation
with probability proportional to their corresponding matrix elements:
\beq
P\left({\rm single}|p,q,r\right) \propto \left|H(r \leftarrow p)\right|, \quad {\rm and} \quad
P\left({\rm double}|p,q,r\right) \propto H_{rpq}^{{\rm tot}},
\eeq
where
\beq
H_{rpq}^{{\rm tot}}\equiv\sum_{s'\notin\left\{ p,q,r\right\} }\left|H(r s' \leftarrow p q)\right|
\label{h_rpq_tot}
\eeq
is the sum of double-excitation off-diagonal elements in which electrons
in spin-orbitals $p$ and $q$ excite to two \emph{other} spin-orbitals, one of which is $r$.
Note that $H_{rpq}^{{\rm tot}}$
can be precomputed and stored at the beginning of the run.

One method of choosing between single and double excitations is to choose a single excitation with probability
$\frac{\left|H(r \leftarrow p)\right|}{H_{rpq}^{\rm tot}+\left|H(r \leftarrow p)\right|}$ and to choose a double excitation
otherwise.
However, the problem with this approach is that if $\left|H(r \leftarrow p)\right| \gg H_{rpq}^{\rm tot}$
the double excitation probability becomes small, so when a double excitation
is chosen, the newly-spawned walker gets a large weight.  A solution to this problem is to make
both a double-excitation and a single-excitation move if
$\left|H(r \leftarrow p)\right| > H_{rpq}^{\rm tot}$.
Using this approach, the probabilities of selecting a single or double excitation are
\beq
\tilde{P}\left({\rm single}|p,q,r\right)=\begin{cases}
\frac{\left|H(r \leftarrow p)\right|}{H_{rpq}^{{\rm tot}}+\left|H(r \leftarrow p)\right|},
& \mbox{ if } \left|H(r \leftarrow p)\right|<H_{rpq}^{{\rm tot}};\\%& \mbox{ if } \frac{\left|H(r \leftarrow p)\right|}{H_{rpq}^{{\rm tot}}+\left|H(r \leftarrow p)\right|} < \frac{1}{2} \mbox{;}\\
1, & \mbox{ otherwise,}
\end{cases}
\eeq
and
\beq
\tilde{P}\left({\rm double}|p,q,r\right)=\begin{cases}
\frac{H_{rpq}^{{\rm tot}}}{H_{rpq}^{{\rm tot}}+\left|H(r \leftarrow p)\right|},
& \mbox{ if } \left|H(r \leftarrow p)\right|<H_{rpq}^{{\rm tot}};\\%\frac{H_{rpq}^{{\rm tot}}}{H_{rpq}^{{\rm tot}}+\left|H(r \leftarrow p)\right|} > \frac{1}{2} \mbox{;}\\
1, & \mbox{ otherwise,}
\end{cases}
\eeq
respectively.  Now, there can still be some large-weight single-excitation moves when $\left|H(r \leftarrow p)\right| \gg H_{rpq}^{\rm tot}$,
but since single-excitation moves are a very small fraction of all moves, this is not a serious problem.

\subsection{Check for correctness of heat-bath sampling}
While all valid double excitations can be proposed using this algorithm, one may wonder whether
the same is true of single excitations.
The only situation in which a valid single excitation $\left(r \leftarrow p\right)$ cannot be proposed by our heat-bath algorithm is
if all other occupied spin-orbitals $q\ne p$ are the only ones of their irreducible representation, since then there would be no symmetry-allowed
double excitation $\left(r s \leftarrow p q\right)$, and $\tilde{P}\left(r|p,q\right)$ would be zero.
Hence, a sufficient $-$ but not necessary $-$ condition for heat-bath sampling to be correct for a given system consisting of $n_\uparrow$ spin-up electrons and $n_\downarrow$ spin-down electrons is:
\beq
\max\left(n_{\uparrow},n_{\downarrow}\right) &>& N_{\rm irrep.}',
\eeq
where $N_{\rm irrep.}'$ is the number of irreducible representations consisting of only one spatial orbital.

Care must be taken that the full symmetry is used in this step.
It is common practice to use $D_{2h}$ symmetry, rather than the full $D_{\infty h}$ symmetry, for a
homonuclear diatomic molecule.
For example, suppose there is only one orbital pair that transforms as the $\Delta_g$ irreducible representation of $D_{\infty h}$.
The member of the pair that transforms as $x^2-y^2$ belongs to the $A_g$ irreducible representation of $D_{2h}$
whereas the member that that transforms as $xy$ belongs to the $B_{1g}$ irreducible representation of $D_{2h}$.
However, there are other functions, 1, $x^2+y^2$ and $z^2$ that also transform as the $A_g$ representation.
So, using the irreducible representations of $D_{\infty h}$ we would conclude that this orbital pair contributes two to $N_{\rm irrep.}'$
whereas using the irreducible representations of $D_{2h}$ symmetry we would incorrectly conclude that
this orbital pair contributes one to $N_{\rm irrep.}'$.

In practice, when we do not know the irreducible representations of the orbitals in the full symmetry group,
we calculate $N_{\rm irrep.}'$ as follows. Consider the system consisting of the
same set of $M$ spatial orbitals but only one electron.
Compute the Full CI Hamiltonian matrix $H'$ (dimension $M$) for this system. The number of columns of $H'$
that have no nonzero off-diagonal elements is equal to $N_{\rm irrep.}'$.
This check is done at the beginning of a run to ensure that heat-bath sampling is unbiased
for the system being examined.

\section{Results: B$_2$, N$_2$ and F$_2$ molecules}
\label{results}

The relative efficiency of the approximate heat-bath method to the uniform method
was tested on the B$_2$, N$_2$ and F$_2$ molecules.
Dunning's cc-pVQZ basis was used for all three molecules, and in addition, for N$_2$,
cc-pV$X$Z bases with $X\in\left\{ {\rm D},{\rm T},{\rm Q},5\right\}$ were used to
study the basis set dependence.
For each basis set, the determinants in both the deterministic space and the trial wavefunction
were selected from the set of determinants that are at most quadruple excitations
from the Hartree-Fock determinant. The size of the deterministic space was
$4\times 10^4$ determinants, and the trial wavefunction contained
$10^3$ determinants (except for cc-pV5Z, which had a trial wavefunction with
only $500$ determinants). To accelerate convergence, time-reversal symmetry (see Appendix~\ref{time_sym})
and natural orbitals from a second-order Moller-Plesset perturbation theory (MP2) calculation
were used.

\begin{figure}
\input{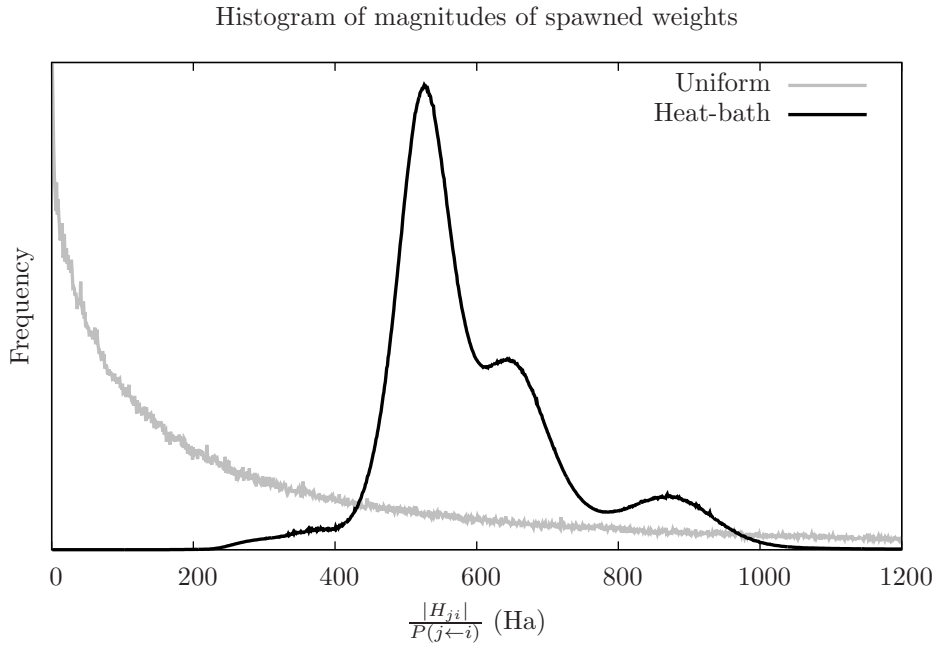}
\protect\caption{Probability distributions of
off-diagonal absolute weights (in multiples of $\tau$),
spawned by the uniform and heat-bath algorithms
for equilibrium ${\rm N}_{2}$ in a cc-pV5Z basis at with all electrons excited and $10^{6}$ walkers. In the
new heat-bath sampling algorithm, the spawned off-diagonal weights
are close in magnitude, whereas in uniform sampling, the
values of the spawned weights span many orders of magnitude.
The location of the peak in the heat-bath distribution corresponds to an
off-diagonal weight of about 0.1.}
\label{wt_frequency_v5z}
\end{figure}

Fig.~\ref{wt_frequency_v5z} shows the frequency of spawned absolute weights (in units of $\tau$) for the
uniform and the approximate heat-bath methods applied to the N$_2$ molecule at equilibrium geometry
in a cc-pV5Z basis.
The vast majority of the absolute weights spawned by the heat-bath method lie in the range 250-1000 Ha
times $\tau$, whereas those spawned by the uniform method range over orders of magnitude.

The reduction in the spread of the spawned weights
makes the initiator approximation~\cite{CleBooAla-JCP-10} more meaningful and
allows one to use a larger time step $\tau$.
In the uniform method, a state can become an initiator simply because
it happened to receive a large weight from a single determinant.
The increase in $\tau$ speeds up the time needed to achieve equilibration, as well as increasing the
efficiency of the method.
Since the statistical uncertainty, $\sigma_{E}$, in the estimate of the ground state energy decreases as
the inverse square root of the Monte Carlo run time $T$, the efficiency is defined to be
\beq
{\rm Efficiency}\propto \frac{1}{\sigma_{E}^2 T}.
\eeq
We note that unlike the diffusion Monte Carlo method where the equilibration time is a negligible part
of the total time, in S-FCIQMC the equilibration can be a substantial part of the total.

Table~\ref{eff_vs_nelec} shows these quantities for a sequence of three molecules.
The efficiency gain increases with the number of electrons from 3.8 for B$_2$, to 54 for F$_2$.
For uniform sampling, when $N$ or $M$ is large, it is hard to get an accurate estimate of the
statistical error because (1) the optimal $\tau$ is very small, making the autocorrelation time in units of
the number of Monte Carlo steps very long, and (2) the population must be sufficiently large to
enable cancellations to occur. Thus, in Tables~\ref{eff_vs_nelec} and~\ref{eff_vs_basis},
the equilibration speedup and increase
in optimal time step are given to only single-digit accuracy.

\begin{table}[htbp]
\begin{tabular}{|c|c|c|c|}
\hline
molecule & efficiency gain & equilibration speedup & $\tau_{{\rm opt}}$ increase\tabularnewline
\hline
\hline
B$_2$ & 3.8 & 5 & 20\tabularnewline
\hline
N$_2$ & 31 & 30 & 200\tabularnewline
\hline
F$_2$ & 54 & 50 & 200\tabularnewline
\hline
\end{tabular}
\caption{Efficiency gains for all-electron equilibrium calculations of the ground state energy in a cc-pVQZ basis for different molecules.
The efficiency gain increases with the number of electrons.
}
\label{eff_vs_nelec}
\end{table}

\begin{figure}
\input{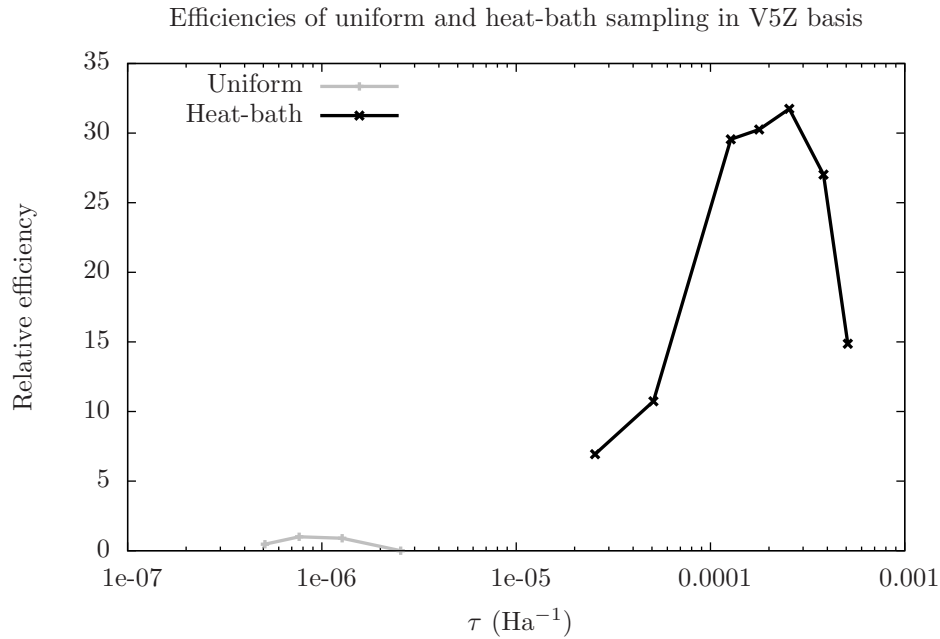}
\protect\caption{Comparison of relative efficiency vs. time step size for uniform and
heat-bath methods on the all-electron nitrogen dimer at equilibrium in the cc-pV5Z basis. Efficiency plotted relative to the greatest
efficiency seen for uniform. Heat-bath
sampling improves efficiency by a factor of 32, while
the optimal time step is about a factor of 300 larger.
}
\label{efficiency_N2_v5z}
\end{figure}

Table~\ref{eff_vs_basis} demonstrates that the efficiency gain increases with the size of the
basis.  For the N$_2$ molecule, the gain increases from 3.4 for the cc-pVDZ basis to
32 for the cc-pV5Z basis.
For the cc-pV5Z basis, Fig.~\ref{efficiency_N2_v5z} shows how the relative efficiencies change
with $\tau$.  The relative efficiencies are normalized such that the peak for uniform sampling
is at one.
Not only is the peak efficiency 32 times higher for heat-bath sampling, but
the peak is broader, which makes it easier to choose a value of $\tau$ that is close to optimal.

\begin{table}[htbp]
\begin{tabular}{|c|c|c|c|}
\hline
basis& efficiency gain & equilibration speedup & $\tau_{{\rm opt}}$ increase\tabularnewline
\hline
\hline
cc-pVDZ & 3.4 & 2 & 5\tabularnewline
\hline
cc-pVTZ & 16 & 30 & 200\tabularnewline
\hline
cc-pVQZ & 31 & 30 & 200\tabularnewline
\hline
cc-pV5Z & 32 & 60 & 300\tabularnewline
\hline
\end{tabular}
\caption{Efficiency gains and the factor by which the optimal time step $\tau_{{\rm opt}}$ increases upon
using heat-bath sampling instead of uniform sampling, for N$_2$ at the equilibrium geometry including core excitations.
For a given sampling algorithm, $\tau_{{\rm opt}}$ is the time step that maximizes the efficiency.
The gain in efficiency of heat-bath sampling over uniform sampling increases with increasing basis set size.
}
\label{eff_vs_basis}
\end{table}

In summary, the gain in efficiency from using heat-bath sampling rather than uniform sampling
increases both with the number of electrons and the number of spin-orbitals.

\section{Discussion}

The approximate heat-bath sampling algorithm introduced in this paper
enables one to sample many-body quantum states connected to an
initial state with probability approximately proportional
to an arbitrary function of the Hamiltonian matrix element
connecting the two states. We chose the
probabilities to be proportional to the corresponding matrix element,
which greatly improves the efficiency of S-FCIQMC.
However, other choices are possible; e.g., in the
Monte Carlo Configuration Interaction method~\cite{greer1998monte},
it would be useful to choose the probabilities to be proportional
to the lowering in energy estimated from second-order perturbation theory.

Finally, we note that the gain in efficiency from using the heat-bath method
will be even greater for excited states if they are calculated using a projector
that involves sampling the square of the Hamiltonian~\cite{BooCha-JCP-12},
since products of weights fluctuate even more than the weights themselves.

{\it Acknowledgements:} We thank Frank Petruzielo, Alessandro Roggero, Matt Otten, and George Booth
for contributing to the development of our S-FCIQMC code.
This work was supported by grants NSF CHE-1112097, DOE DE-SC0006650, and NSF ACI-1534965.
After moving to UIUC, H.J.C. was supported by the SciDAC program under DOE Award Number DE-FG02-12ER46875.

\appendix
\section{Recap of algorithm when applied to S-FCIQMC}
\label{Algorithm}

Before the S-FCIQMC run, compute and store the quantities $S_p$, $D_{pq}$,
$\tilde{P}\left(r|p,q\right)$, $\tilde{P}\left(s|p,q,r\right)$, and $H_{rpq}^{{\rm tot}}$,
as described in Appendix~\ref{Precompute}.

When spawning off-diagonal moves from determinant $\left|i\right>$, first
compute and store the probability distribution for selecting the first occupied spin-orbital $p$ from this determinant,
\beq
\tilde{P}\left(p\right)=\frac{S_p}{\sum_{p'}S_{p'}},
\eeq
and compute the $A$ and $Q$ arrays needed for the alias method (see Appendix~\ref{Alias}).
This step takes $\mathcal{O}\left(N\right)$ time.

Divide up the weight $w_i$ on $\left|i\right>$ into $n_i=\max\left(1,\lfloor \left| w_i \right|\rceil\right)$ walkers.
Then, for each walker on determinant $\left|i\right>$, spawn one or two new determinants as follows:
\begin{enumerate}

\item Using the alias method, choose the first occupied spin-orbital $p$ from the stored distribution
$\tilde{P}\left(p\right)$. This step takes $\mathcal{O}\left(1\right)$ time.

\item Compute the probability distribution for choosing the second occupied spin-orbital $q$ given that $p$ was
already selected,
\beq
\tilde{P}\left(q|p\right)=\frac{D_{pq}}{\sum_{q'}D_{pq'}},
\eeq
and sample the second occupied spin-orbital $q$ from $\tilde{P}\left(q|p\right)$. This step takes $\mathcal{O}\left(N\right)$ time.

\item Choose a spin-orbital $r\notin \left\{p,q\right\}$ from the stored distribution
$\tilde{P}\left(r|p,q\right)$,
such that $p$ and $r$ are of the same spin, using the alias method.
If $r$ is occupied in $\left|i\right>$, no moves are generated by this walker; go to next walker. Otherwise, $r$ is the first
unoccupied spin-orbital for this excitation. This step takes $\mathcal{O}\left(1\right)$ time.

\item Decide whether to spawn a single excitation, a double excitation or one of each, as follows. Compute the
single-excitation matrix element $H\left(r \leftarrow p\right)$. If $|H\left(r \leftarrow p\right)|$ is greater than the stored
quantity $H_{rpq}^{{\rm tot}}$, then spawn both the single excitation $\left(r \leftarrow p\right)$
and a double excitation; otherwise spawn only one of the two types of excitations as follows. With probability
\beq
\frac{|H\left(r \leftarrow p\right)|}{H_{rpq}^{{\rm tot}}+|H\left(r \leftarrow p\right)|},
\eeq
choose the single excitation $\left(r \leftarrow p\right)$; otherwise choose a double excitation. This step takes $\mathcal{O}\left(N\right)$ time,
since that is the complexity of computing a single-excitation matrix element.

\item If a double excitation is to be proposed, choose a fourth spin-orbital $s\notin \left\{p,q,r\right\}$ from the stored distribution
$\tilde{P}\left(s|p,q,r\right)$,
using the alias method.
If $s$ is occupied in $\left|i\right>$, no double excitation is generated by this walker.
Otherwise, $s$ is the second unoccupied spin-orbital for this excitation, and the move $\left(r s \leftarrow p q\right)$
is generated. This step takes $\mathcal{O}\left(1\right)$ time.

\item After generating the excitation(s), compute the weight spawned on the new determinant(s) as follows.
If a single excitation
$\left(r \leftarrow p\right)$ was proposed to a new determinant $\left|j\right>$, the weight spawned on $\left|j\right>$ is
\beq
w_{j}^{\left(t+1\right)} &=&\frac{-\tau H\left(r \leftarrow p\right)}{P\left(r \leftarrow p\right)}\left(\frac{w_{i}^{t}}{n_i}\right),
\eeq
where $P\left(r \leftarrow p\right)$ is given in Appendix~\ref{ProposalProbs}.
If a double excitation
$\left(r s \leftarrow p q\right)$ was proposed to a new determinant $\left|k\right>$, the weight spawned on $\left|k\right>$ is
\beq
w_{k}^{\left(t+1\right)} &=&\frac{-\tau H\left(r s \leftarrow p q\right)}{P\left(r s \leftarrow p q\right)}\left(\frac{w_{i}^{t}}{n_i}\right),
\eeq
where $P\left(r s \leftarrow p q\right)$ is given in Appendix~\ref{ProposalProbs}.

\end{enumerate}

\section{Proposal probabilities}
\label{ProposalProbs}

As discussed in the text, when a new walker is spawned, the weight assigned to that walker
is proportional to the corresponding matrix element of the projector divided by the proposal probability
for that transition.  The computation of these proposal probabilities is described next.
The proposal probabilities both for single and
double excitations can be computed in $\mathcal{O}\left(N\right)$ time.

\subsection{Single excitations}

The probability for a single excitation $\left(r \leftarrow p\right)$ is
\beq
P\left(r \leftarrow p\right) &=& \sum_{q'\in {\rm occ.}}\tilde{P}\left(p\right)\tilde{P}\left(q'|p\right)\tilde{P}\left(r|p,q'\right)
\tilde{P}\left({\rm single}|p,q,r\right),
\label{single_excit_prob}
\eeq
where
\beq
P\left({\rm single}|p,q,r\right)=\begin{cases}
\frac{\left|H(r \leftarrow p)\right|}{H_{rpq}^{{\rm tot}}+\left|H(r \leftarrow p)\right|},
& \mbox{ if } \left|H(r \leftarrow p)\right|<H_{rpq}^{{\rm tot}};\\%\frac{\left|H(r \leftarrow p)\right|}{H_{rpq}^{{\rm tot}}+\left|H(r \leftarrow p)\right|} < \frac{1}{2} \mbox{;}\\
1, & \mbox{ otherwise,}
\end{cases}
\eeq
is the probability of choosing the single excitation $\left(r \leftarrow p\right)$ given that
spin-orbitals $\left\{p,q,r\right\}$ have already been selected.

An alternative probability is
$P'\left(r \leftarrow p\right)=\left(N-1\right)\tilde{P}\left(p\right)\tilde{P}\left(q|p\right)\tilde{P}\left(r|p,q\right)$, where $q$ is the spin-orbital that was selected and discarded during the
heat-bath sampling routine. As described in Appendix~\ref{Multiple_events}, this alternative
method has an efficiency tradeoff associated with it:
it increases efficiency by avoiding having to compute all terms in the sum, but it also
decreases efficiency by increasing the fluctuations in spawned weights. This alternative probability
was not employed in this paper.

\subsection{Double excitations}

The probability for a double excitation $\left(r s \leftarrow p q\right)$ is
\beq
P\left(r s \leftarrow p q\right) &=& \tilde{P}\left(p\right)\tilde{P}\left(q|p\right) \left[
     \tilde{P}\left(r|p,q\right)
\tilde{P}\left({\rm double}|p,q,r\right)\tilde{P}\left(s|p,q,r\right)
+\tilde{P}\left(s|p,q\right)
\tilde{P}\left({\rm double}|p,q,s\right)\tilde{P}\left(r|p,q,s\right)
 \right] \nonumber \\
&& +\;\tilde{P}\left(q\right)\tilde{P}\left(p|q\right) \left[  \tilde{P}\left(r|q,p\right)
\tilde{P}\left({\rm double}|q,p,r\right)\tilde{P}\left(s|q,p,r\right)
+\tilde{P}\left(s|q,p\right)
\tilde{P}\left({\rm double}|q,p,s\right)\tilde{P}\left(r|q,p,s\right)
\right], \nonumber \\
\label{double_excit_prob}
\eeq
where
\beq
\tilde{P}\left({\rm double}|p,q,r\right)=\begin{cases}
\frac{H_{rpq}^{{\rm tot}}}{H_{rpq}^{{\rm tot}}+\left|H(r \leftarrow p)\right|},& \mbox{ if } \left|H(r \leftarrow p)\right|<H_{rpq}^{{\rm tot}};\\%\frac{H_{rpq}^{{\rm tot}}}{H_{rpq}^{{\rm tot}}+\left|H(r \leftarrow p)\right|} > \frac{1}{2} \mbox{;}\\
1, & \mbox{ otherwise,}
\end{cases}
\eeq
is the probability of choosing a double excitation given that $\left\{p,q,r\right\}$ have already been selected.

The second and third terms in Eq.~\ref{double_excit_prob} are zero for opposite-spin excitations.

As with single excitations, an alternative probability is
\beq
P'\left(r s \leftarrow p q\right) = c\tilde{P}\left(p\right)\tilde{P}\left(q|p\right)\tilde{P}\left(r|p,q\right)\tilde{P}\left({\rm double}|p,q,r\right)\tilde{P}\left(s|p,q,r\right)\tilde{P}\left(p\right)\tilde{P}\left(q|p\right)\tilde{P}\left(r|p,q\right),
\eeq
where $c$ is either 2 or 4 for opposite-spin or same-spin double excitations, respectively.
Again, this alternative probability was not employed in this paper.

\section{Quantities that must be precomputed at start of run}
\label{Precompute}

At the start of the run, compute and store the following tensors.
It is assumed that up- and down-spin orbitals are the same, but the number of
up- and down-spin electrons need not be the same.
\begin{enumerate}

\item Electron pair selection probabilities tensor:
\begin{eqnarray}
D_{pq} & = & \sum_{r's'}\left|H(r' s' \leftarrow p q)\right|%+\sum_r \left<\left|H\left(r \leftarrow p\right)\right|\right>
\end{eqnarray}
for all $p\ne q$. If single excitations were ignored, this would
roughly correspond to the relative heat-bath probability of selecting
a pair of electrons to excite, i.e., $\tilde{P}\left(p,q\right)\propto D_{pq}$. This has
size $2M(2M-1)/2$, since electrons can be of either spin.
The storage could be reduced to $M(M-1)/2$ for same-spin electrons and
and $M^2$ for opposite-spin electrons.

\item Single electron selection probabilities tensor:
\beq
S_{p} &=& \sum_{q'\ne p}D_{pq'}
\eeq
for all $p$. The electrons are chosen one at a time with probabilities $\tilde{P}\left(p\right)$
and $\tilde{P}\left(q|p\right)$ computed using $S_p$ and $D_{pq}$, respectively. $S$ has size $M$.

\item First hole selection probabilities tensor:
\beq
\tilde{P}\left(r|p,q\right) &=& \frac{\sum_{s'}\left|H(r s' \leftarrow p q)\right|}%+\left<\left|H\left(r \leftarrow p\right)\right|\right>}
{\sum_{r's'}\left|H(r' s' \leftarrow p q)\right|}
\eeq
for all $p\ne q\ne r$. This is the probability of choosing the first empty
spin-orbital $r$, given that electrons in spin-orbitals $p$ and $q$ have already
been selected for excitations. This can be stored as two separate
tensors, $\tilde{P}_{\rm same}\left(r|p,q\right)$ and $\tilde{P}_{\rm opposite}\left(r|p,q\right)$,
for same-spin and opposite-spin double excitations, respectively.
The same-spin excitation tensor only has to be of size $\frac{1}{2}M^{3}$,
while the opposite-spin excitation tensor can be of size $M^{3}$.
In that case, $\tilde{P}_{\rm opposite}\left(r|p,q\right)$ represents
the probability of choosing $r$ given $p,q$ with $r$ and $p$ having
same spin.

\item Double excitations probabilities tensor:
\beq
\tilde{P}\left(s|p,q,r\right) &=& \frac{\left|H(r s \leftarrow p q)\right|}{\sum_{s'}\left|H(r s' \leftarrow p q)\right|}
\eeq
for all $p\ne q\ne r\ne s$. When sampling a double excitation, this is the probability of choosing
the second empty spin-orbital $s$ given that electrons in spin-orbitals $p$ and $q$ and empty spin-orbital $r$ have already been chosen.
Also, store the denominators $H_{rpq}^{{\rm tot}}=\sum_{s'}\left|H(r s' \leftarrow p q)\right|$,
the summed magnitudes of double excitation matrix elements. This will
be needed for comparing with single excitation matrix elements.
The tensor can be stored as two separate tensors, $\tilde{P}_{\rm same}\left(s|p,q,r\right)$
and $\tilde{P}_{\rm opposite}\left(s|p,q,r\right)$, of sizes $\frac{1}{2}M^{4}$
and $M^{4}$, respectively.
\end{enumerate}
For each of the two probability tensors $\tilde{P}\left(r|p,q\right)$ and $\tilde{P}\left(s|p,q,r\right)$,
we store not only the probabilities,
but the corresponding $A$ and $Q$ tensors for sampling them in $\mathcal{O}\left(1\right)$
time using the alias method (see Appendix~\ref{Alias}). Thus, the total storage space for $\tilde{P}\left(s|p,q,r\right)$
and its corresponding $A$ and $Q$ tensors is $\frac{3}{2}M^{4}$
integers and $3M^{4}$ single-precision real numbers. In comparison,
the integrals files that are already being stored are $\mathcal{O}\left(\frac{1}{8}M^{4}\right)$
double-precision real numbers, so the storage requirement is 18 times larger than the storage
requirement for the integrals alone.

\subsection{Relaxing the storage requirements}
\label{reduce_storage}

While the above storage requirements have the same $\mathcal{O}\left(M^4\right)$ scaling as the
two-body integrals that are already stored, the fact that $\tilde{P}\left(s|p,q,r\right)$
requires 18 times the storage of the integrals can become prohibitive for more than about
250 orbitals.
One possibility is to make a Cauchy-Schwarz approximation for estimating the 2-body
integrals~\cite{SmaBooAla-ARX-15}.
Instead, here we discuss a method that avoids storing $\tilde{P}\left(s|p,q,r\right)$
for all $\mathcal{O}\left(M^4\right)$ combinations $\left\{p,q,r,s\right\}$
without making the Cauchy-Schwarz approximation.

First, note that $\tilde{P}\left(s|p,q,r\right)$ is only sampled when $\left\{p,q,r\right\}$ have already
been selected. Some $\left\{p,q,r\right\}$ combinations occur a lot more often than others, so
$\tilde{P}\left(s|p,q,r\right)$ needs to be stored only for the most frequently-occurring triplets
$\left\{p,q,r\right\}$. For the other, less frequently-occurring triplets $\left\{p,q,r\right\}$, one
could either choose spin-orbital $s$ uniformly, or compute the correct heat-bath
conditional probability $\tilde{P}\left(s|p,q,r\right)$ on the fly in $\mathcal{O}\left(M\right)$ time.

\begin{figure}[h]
\input{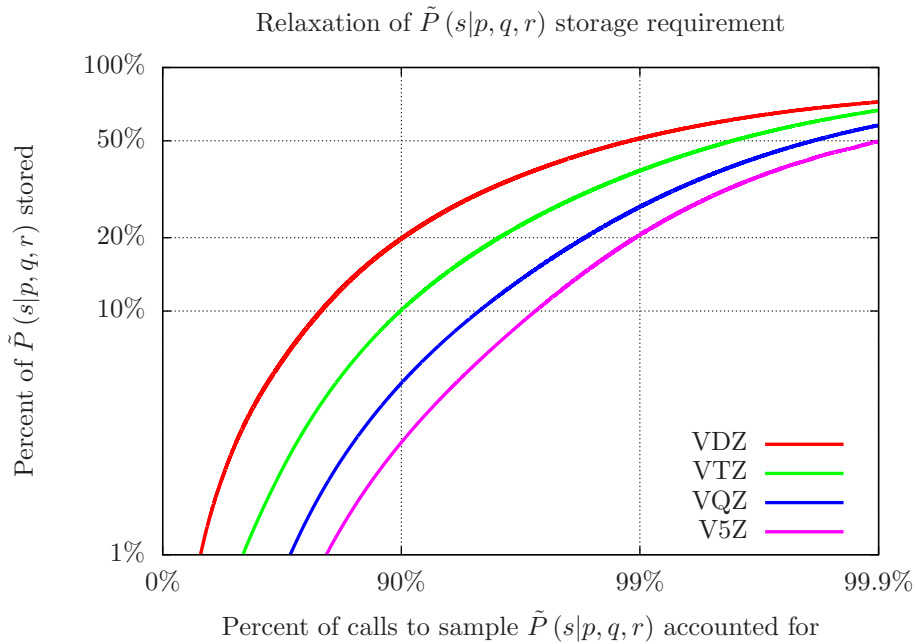}
\protect\caption{
This plot shows what percentage of $\tilde{P}\left(s|p,q,r\right)$ must be stored,
in order for the stored part of $\tilde{P}\left(s|p,q,r\right)$ to account for a
given percentage of the total calls to sample it,
for an all-electron nitrogen dimer
in cc-pV$X$Z basis sets for $X\in\left\{D,T,Q,5\right\}$.
It was obtained by counting the number of times that each triplet $\left\{p,q,r\right\}$ was selected (followed
by sampling $\tilde{P}\left(s|p,q,r\right)$)
during a run with $10^5$ walkers.
}
\label{pqr_count}
\end{figure}

As can be seen in Fig.~\ref{pqr_count}, the fraction of $\left\{p,q,r\right\}$ triplets that account for
the vast majority of calls to sample $\tilde{P}\left(s|p,q,r\right)$ decreases with basis set size.
In order to account for 99\% of the number of calls to sample $\tilde{P}\left(s|p,q,r\right)$,
for cc-pVDZ, one must store $\tilde{P}\left(s|p,q,r\right)$ for
about 50\% of the $\left\{p,q,r\right\}$ triplets,
while for cc-pV5Z, it is only required to store $\tilde{P}\left(s|p,q,r\right)$ for
about 20\% of the triplets.

The largest basis used in this paper is cc-pV5Z (182 orbitals), but with this modification,
calculations with a cc-pV6Z basis (280 orbitals) could easily be done.

\section{Alias method for sampling from discrete distributions}
\label{Alias}
Discrete distributions consisting of $M$ probabilities can be sampled straightforwardly by constructing an array of cumulative
probabilities, drawing a random number, and doing a binary search to find the interval in which
the random number falls.  This requires $\mathcal{O}\left(M\right)$ time to set up the cumulative probabilities
and ${\mathcal O}(\log M)$ time to sample.

However, when the same distribution is being sampled repeatedly, the alias method~\cite{walker1977efficient}
can be used to sample it in even more quickly, in ${\mathcal O}(1)$ time.
The method employs a real array, $Q$, and an integer array $A$, each of length $M$.
Sampling with the alias method is done as follows.
First, orbital $i$ is selected uniformly from the set of all $M$ possible orbitals. Then, with probability $Q_i$,
orbital $i$ is sampled; else, orbital $A_i$ is sampled.
The cost of doing this is just the cost of drawing two uniform random numbers.

The set of probabilities $\left\{Q_i\right\}$ and aliases $\left\{A_i\right\}$
can be generated in $\mathcal{O}\left(M\right)$ time at the beginning of the run~\cite{kronmal1979alias}
as shown in the pseudocode in Fig.~\ref{alias_setup}.

\begin{figure}[h]
\protect\caption{Alias method setup}\label{alias_setup}
\begin{framed}
\begin{verbatim}
! Inputs:  M = number of discrete states to sample
!          {P(i)} = discrete set of probabilities of sampling state i
!
! Outputs: {A(i)} = aliases of states
!          {Q(i)} = probabilities of returning state i rather than its alias A(i)

! Scale probabilities by M and place in two lists, those smaller and those bigger than 1.
n_s = 0 ; n_b = 0
do i=1,M
  A(i) = i           ! Not an arbitrary initialization (*)
  Q(i) = M*P(i)
  if (Q(i)<1) then
    n_s = n_s + 1
    smaller(n_s) = i
  else
    n_b = n_b + 1
    bigger(n_b) = i
  endif
enddo

! For each orbital construct probability of staying on the orbital and the alias of the orbital.
do while (n_s > 0 .and. n_b > 0)
  s = smaller(n_s)
  b = bigger(n_b)
  A(s) = b
  Q(b) = Q(b) + Q(s) - 1
  if (Q(b) < 1) then
    smaller(n_s) = b
    n_b = n_b - 1
  else
    n_s = n_s - 1
  endif
enddo

! (*) Initialize A(i) = i, so that if the random number exceeds a floating point approximation
!     to Q(i) = 1, i is returned (as it should be) rather than an arbitrary initialized state.
\end{verbatim}
\end{framed}
\end{figure}

\section{Time-reversal symmetry}
\label{time_sym}

The size of the Hilbert space can be reduced by almost a factor of two when the
number of up and down electrons are equal by taking advantage of time-reversal symmetry.
This increases the effectiveness of cancellations, and allows one to use
larger deterministic spaces and trial wavefunctions in S-FCIQMC.
It also enables one to calculate an excited state as easily as the ground state provided
that it is the lowest state of a different symmetry under time reversal than the ground state;
e.g., if the ground state is a singlet, it enables calculating the lowest triplet state.
All calculations in this paper make use of time-reversal symmetry
and here we discuss details of how it is implemented in S-FCIQMC.

Let $\hat{T}$ denote the time-reversal operator, so $|i'\rangle \equiv \hat{T} |i\rangle $
is the state obtained by flipping all the spins of the electrons
in state $|i\rangle$. Since $\hat{T}^2 |i\rangle = |i\rangle $,
the eigenvalues of the time-reversal operator, $z$, must be $\pm 1$.
Even $S$ states have $z=1$, and odd $S$ states have $z=-1$, where $S$ is the total spin of the system.
So, the wavefunctions can be expanded in a symmetrized basis,
\beq
\label{eq:sym}
|\tilde{i}\rangle= \frac{1}{\sqrt{2}N_i}\left( |i\rangle + z |i'\rangle \right)
\eeq
where,
\beq
N_i=\begin{cases}
1,&\mbox{ if } \left<i|i'\right>=0,\\
\sqrt{2}, & \mbox{ otherwise },
\end{cases}
\eeq
is a normalization factor. Note that a state which is its own time-reversed partner ($|i\rangle = |i'\rangle$)
can only have non-zero coefficients in a wavefunction that has $z=1$.

In each symmetrized linear combination, $|\tilde{i}\rangle$, one of the states is
chosen to be the ``representative", $|i\rangle_{\rm rep}$.  The representative
has a positive coefficient in the linear combination, and non-representative state has the same coefficient
multiplied by $z$.  In Eq.~\eqref{eq:sym} $|i\rangle_{\rm rep} = |i\rangle$.
The representative is chosen by converting the computer
representation of states (a binary string) into a number and then defining,
\begin{equation}
|i \rangle_{\rm rep} \equiv \min \left( |i\rangle, |i'\rangle \right)
\end{equation}
Thus the representative is the state that comes first
according to an arbitrary (but consistent) convention for ordering.

Recall that when time-reversal symmetry is not used,
a walker that makes an off-diagonal move from state $|i \rangle $ to state $| j  \rangle$ is
assigned weight, $w_j=\frac{-\tau H_{ji}}{P_{ji}}\left(\frac{w_i}{n_i}\right)$,
where $w_i$ and $n_i$ are the weight and number of walkers on $|i\rangle$, respectively.
This viewpoint is adopted even with the inclusion of time-reversal symmetry, albeit with
modifications to the Hamiltonian matrix element and the proposal probability, to give,
\begin{equation}
	w_{\tilde{j}} = -\frac{\tau \tilde{H}_{\tilde{j} \tilde{i}}} {P_{\tilde{j}\tilde{i}}} \left( \frac{w_{\tilde{i}}}{n_{\tilde{i}}}\right)
\end{equation}
where $\tilde{H}_{\tilde{j} \tilde{i}}$ is the Hamiltonian matrix element between
the two symmetrized pairs of states, $\tilde{i}$ and $\tilde{j}$, and $P_{\tilde{j}\tilde{i}}$
is the total probability of making a transition between them.

Since the time-reversal operator and the Hamiltonian commute ($[\hat{T},\hat{H}]=0$), it follows that $H_{ji'}=H_{j'i}$,
and since $\hat{T}^{2} = 1$, $H_{j'i'}=H_{ji}$.
Then, using expressions for $| \tilde{i} \rangle $ and $| \tilde{j} \rangle$ from Eq.~\eqref{eq:sym} we get,
\begin{equation}
	\tilde{H}_{\tilde{j} \tilde{i}} = \frac{ H_{ji} + z H_{j'i}} {N_i N_j}
\label{eq:sym_ham}
\end{equation}

To evaluate $P_{\tilde{j}\tilde{i}}$, treat $|i \rangle_{\rm rep}$ just as the
usual (unsymmetrized) state (only consider the usual Hamiltonian connections from
$|i \rangle_{\rm rep}$, not its symmetry related pair state).
With probability $P_{j,i_{\rm rep}}$, state $|i \rangle_{\rm rep}$
spawns to state $|j\rangle$, which may or may not be its own representative.
Since non-representative states are not included in the list of occupied states,
if $|j'\rangle \ne |j\rangle$, we must sum over both possibilities to get
\begin{eqnarray}
	P_{\tilde{j}\tilde{i}} &=& P_{j_{\rm rep},i_{\rm rep}}  \nonumber \\
							&=& P_{j,i_{\rm rep}} + P_{j',i_{\rm rep}}.
\end{eqnarray}

\section{Choice of weighting when multiple events lead to a single state}
\label{Multiple_events}
Let $P$ be the matrix element of the projector for a transition between two particular states.
Suppose there are $N$ events with probabilities $p_i, \cdots , p_N$ that result in this transition.
(In general, $\sum_i^N p_i < 1$, since there are other states that can be accessed also.)
The probability of getting that state is $\sum_i^N p_i$, so the weight multiplier for this move
is $\frac{P}{\sum_i^N p_i}$ regardless of which event led to this state.
This prescription is correct since
\beq
\sum_i^N p_i \; \frac{P}{\sum_i^N p_i} = P.
\eeq
($P$ is independent of $i$ because all the possibilities, $i$, correspond to the same state.)

There is an alternative approach which avoids the expense of computing
the probabilities for the $N-1$ other events that result in the same state.
In this approach, when we pick a particular event, $i$, the weight multiplier is $\frac{P}{N p_i}$ rather than $\frac{P}{\sum_i^N p_i}$, so it depends on
which of the $N$ events was selected.
This prescription is also correct since
\beq
\sum_i^N p_i \frac{P}{N p_i} &=& P
\eeq
There is a loss of efficiency, that can be large if the $p_i$ differ greatly from each other.
So, there is a trade-off between avoiding the expense of computing the $N-1$ additional probabilities and
the increase in fluctuations of the weights.

\bibliographystyle{apsrev4-1}
\bibliography{heatbath}

\end{document}